**The impact of the Hall effect on high energy density plasma jets**

P.-A. Gourdain, C. E. Seyler

Abstract

Using a 1-MA, 100ns-rise-time pulsed power generator, radial foil configurations can produce strongly collimated plasma jets. The resulting jets have electron densities on the order of $10^{20}$ cm$^{-3}$, temperatures above 50 eV and plasma velocities on the order of 100 km/s, giving Reynolds numbers of the order of $10^3$, magnetic Reynolds and Péclet numbers on the order of 1. While Hall physics does not dominate jet dynamics due to the large particle density and flow inside, it strongly impacts flows in the jet periphery where plasma density is low. As a result, Hall physics affects indirectly the geometrical shape of the jet and its density profile. The comparison between experiments and numerical simulations demonstrates that the Hall term enhances the jet density when the plasma current flows away from the jet compared to the case where the plasma current flows towards it.

High energy density (HED) plasmas are an exciting research medium to study extreme states of matter. With kinetic pressures larger than one megabar, they open a new range of opportunities in understanding the properties of warm dense matter and the kinematics of flows at large Reynolds (Re = ρvL/μ), magnetic Reynolds (Re$_M$ = v$_A$Lμ$_0$/η) and Péclet (Pe = vL/χ), numbers. In computing Re, Re$_M$, and Pe it is found that the flows in accretion disks surrounding black holes, proto stars and active galaxies are likely to be turbulently advective with significant magnetic fields. When using the mass density ρ, speed v, dynamic viscosity μ, magnetic diffusivity η/μ$_0$, the scale length of the flow L and heat diffusivity χ of plasmas, these numbers scale as: Re ∝ vLZ$^4$A$^{1/2}$n$_i$T$^{-5/2}$, Re$_M$ ∝ vLT$^{3/2}$Z$^{-1}$ and Pe ∝ vLZ(Z+1)n$_i$T$^{-5/2}$. Here n$_i$ is the plasma ion number density, Z the ionization number, A the atomic mass and T the plasma temperature. So, even if scale lengths are small in laboratory experiments, HED plasmas are extremely dense and these numbers can reach large values, placing them at the forefront of laboratory astrophysics[1]. Ultimately only numerical codes can possibly encompass the sizes of most astrophysical objects and the art of numerical simulations does reside in finding the models which can best represent the phenomena observed by astrophysicists. In particular, laboratory experiments can help to validate these numerical codes. While kinetic effects cannot be ignored in astrophysical plasmas, today's large scale simulation efforts focus on the less computationally intensive fluid models. However the magnetohydrodynamics (MHD) model may not capture correctly celestial mechanics. For instance the Hall effect, which is absent from the MHD model, was shown to play a critical role in space dynamics, for instance in the magnetic polarity of galactic jets[2]. This letter wishes to highlight how the Hall term can alter the properties of a strongly collimated HED plasma jet produced by the ablation of a thin metallic foil and it will discuss how these results can be scaled back to astrophysical plasmas. While the jet absolute parameters fall short of astrophysical jets, dimensionless parameters, such as the ratio of the jet radius to its length or the ratio of jet density to the background plasma density, are similar to that of astrophysical jets. While Re, Re$_M$ or Pe are smaller, this letter shows that the importance of the Hall term is actually linked to the ion inertial length, the scale length of the system and the Alvén Mach number rather than Re or Pe. However the impact of large Re and Pe flows on the Hall electric field may reduce its impact.

In the extended magnetohydrodynamics (XMHD) framework, the Hall term generates an electric field perpendicular to the resistive electric field when electrical currents flows across magnetic fields. This Hall electric field can be easily included in Ohm's law to give a simplified version of the generalized Ohm's law (GOL),

$$\boldsymbol{E} = -\boldsymbol{v} \times \boldsymbol{B} + \frac{1}{en_e}\boldsymbol{J} \times \boldsymbol{B} + \eta \boldsymbol{J}. \qquad (1)$$

All bold quantities here represent vectors. $E$ is the total plasma electric field, $v$ the flow velocity, $\eta$ the plasma resistivity, $n_e$ is the electron number density, $J$ the electrical current, $B$ the magnetic field and $e$ is the electron charge. In this equation we have ignored the contribution of the electron pressure to the plasma electric field. The Hall term, $\boldsymbol{J} \times \boldsymbol{B}$, can be neglected in particular situations that we explain now by rewriting the GOL in its dimensionless form, i.e.

$$\boldsymbol{E} = \left(\frac{\delta_i}{L}\boldsymbol{J} - M_A\right) \times \boldsymbol{B} + \frac{1}{S}\boldsymbol{J}. \qquad (2)$$

Eq. (2) was obtained dividing Eq. (1) by the local plasma characteristic velocity of the flow $v_0$ and the local magnetic field $B$. It is important to note that all terms in Eq. (2) are now dimensionless except for the characteristic length scale of the jet $L$ and the ion inertial length $\delta_i$. Their ratio is dimensionless. In general, the ion inertial length $\delta_i = (m_i/\mu_0 e^2 Z^2 n_i)^{1/2}$ is the distance below which ion motion decouples from electron motion, which is still frozen in the magnetic field[3]. We assume here electrical quasi-neutrality, i.e. $n_i=Zn_e$. To keep scaling parameters consistent across all the XMHD equations, the characteristic flow velocity $v_0$ *has to be* the local Alfvén speed $u_A$, i.e. $B/(m_i n_i \mu_0)^{1/2}$. As a result, the dimensionless Eq. (2) uses the Alfvén Mach vector **$M_A$**, which is the ratio of the plasma local velocity *vector* to the local Alfvén speed. In this case the $Re_M$ is also the Lundquist number *S*. Hall physics introduces a great numerical challenge in computing plasma flows on time scales far below the characteristic electron frequencies. MHD codes assume that $\delta_i/L$ is small and simply drop the Hall term from Eq. (2) thereby greatly reducing the total computational time. This letter makes the point that the systematic dismissal of Hall physics based on the presumed smallness of $\delta_i/L$ is shown to be ill-advised in flows with low Alfvén Mach numbers (<<10).

In recent years, we have explored an HED experiment using the plasma produced by a thin metallic foil to test the impact of Hall physics on HED plasma jets. In this experimental setup, the foil is stretched on the anode of a pulsed power generator and connects to the cathode via a hollow metallic pin placed under the foil along the foil geometrical axis. Published research conducted at Cornell University[4,5,6] and Imperial College[7,8,9] presents in greater details the properties and potential applications of such configurations, henceforward called radial foil configurations. The basic idea is that plasma currents converge towards the central pin and JxB forces lift the foil upwards. During this process a small portion of the total plasma current (~ 5 to 10%) flows above the foil where Ohmic resistance heats the plasma. The ablating plasma expands into the vacuum and drags electrical current away from the foil surface. Most of the ablation and plasma motion occurs near the pin, where the JxB forces are intense due to radial convergence. The ablated plasma is forced onto the geometrical foil axis by magnetic pressure and forms a dense, vertical jet with axial velocity on the order of 80 km/s as measured in Ref. 4. The ablated plasma and the base of the jet are visible on the experimental laser Schlieren images presented in Figure 1. Plasma Schlieren imaging[10] records only the light rays which have been diffracted away from the optical focus of the collection optics by electron density gradients. Such regions appear dark in the figure due to publication imperatives. As time progresses, a plasma bubble forms, then expands into the low density plasma above the foil. Kink instabilities[11] disrupt the column at the center of the bubble which quickly breaks apart. While reproducibility of the plasma bubble phase is not guaranteed due to instabilities, the plasma jet phase reproduces nicely from shot to shot as long as current drive waveforms are similar. Using experimental plasma properties previously published[4,5,6] for jet and ablated plasmas, we can estimate the following dimensionless plasma parameters for the jet : Re~$10^3$, $Re_M$ ~ 1, Pe ~ 1 and $M_A$ ~ 10; and in the ablated plasma: Re~$10^4$, $Re_M$~10, Pe~10 and $M_A$~ 2. To highlight the Hall effect experimentally when cathode and anode shapes are different, the current direction has to be reversed. While the plasma velocity stays the same, all electromagnetic terms on the right hand side of Eq. (2) change signs but the Hall term (i.e. **JxB**). A dozen shots were done with standard (radially inward) and reverse (radially outward) electrical currents. We present herein the discharges which highlight best the impact of the Hall effect on plasma dynamics.

Figure 1 shows the differences of the ablated plasma for standard (left) and reverse (right) currents. The plasma instabilities responsible for the elongated diffraction patterns (direction highlighted by the arrows) caused by inhomogeneity in ablation plasma density[5] point away from the axis for standard currents (SC), whereas they point towards the axis for a reverse currents (RC). Since the ablation differs for both plasmas it seems reasonable to assume that the jet density will be affected by the current direction. Indeed laser interferometry shows substantial differences in the jet density profiles. Using a fringe-counting algorithm, such as the IDEA code[12], it is relatively straightforward to obtain the areal electron number density of both jets and their surroundings from laser interferometry (150 ps pulse length at 532 nm). At this wavelength, one fringe shift corresponds to an electron areal number density of $3.72 \times 10^{17}$ cm$^{-2}$. Since the plasma dynamics is quasi-axisymmetric during the early stages of the plasma discharge, it is possible to obtain the local electron number density using a robust Abel inversion technique[13]. Figure 2 shows the experimental local electron density for standard (left panel) and reverse (right panel) currents. Gray masks hide the location where densities could not be computed properly due to the absence of or an inaccuracy in counting interference fringes. Overall, both jets have similar radii, on the order of 400 µm. However, the RC jets are taller. The local electron density varies from $10^{20}$ cm$^{-3}$ at the base of both jets, to $5\times10^{19}$ cm$^{-3}$ at mid height and $1.5\times10^{19}$ cm$^{-3}$ at the top of the jets. Additionally, the RC jet has a larger electron density for a given height as compared to the SC jet. It is important to note that the RC jet interferogram was taken 4 ns earlier than the SC jet. Overall the reverse current case seems to confine the plasma better on axis, thus enhancing the jet density. This effect is easily seen on the plasma electron density profiles presented in Figure 3 at 1.5 mm and 2 mm above the foil. At each height, the plasma densities for both cases were renormalized to highlight profile dissimilarities instead of local density differences. The SC jet profiles are broader than RC profiles. Further SC profiles have a tendency to be flat or slightly hollow near the jet axis. RC profiles are systematically peaked, indicating that the plasma is pushed on axis with greater strength.

Both jets are also visible on data captured by XUV four frame pinhole camera. Due to diffraction caused by the 50 micron pinholes, photon energies below 40 eV are cut-off and hardly reach the photocathode of the quadrant camera, giving a lower bound on the electron temperature of the plasma jet. This energy corresponds to an ionization number of the aluminum plasma on the order of 3. As a result, the ion inertial length $\delta_i$ is on the order of 10 µm, 100 µm and 150 µm at the base, mid height and top of both jets respectively. If we take the 200 µm jet radius as the characteristic length L, the ratio $\delta_i/L$ is smaller than 1 inside the plasma jet and much smaller at the base of the jet. A better characteristic scale length *L* is the magnetic field scale length which reduces to

$$\frac{B}{\|\nabla \times \mathbf{B}\|} = \frac{B}{\mu_0 J} = \frac{\mu_0 I}{2\pi r}\frac{1}{\mu_0 J} \sim \frac{I}{2\pi r}\frac{\pi r^2}{I} = \frac{r}{2} \quad (3)$$

for axi-symmetric systems. Even with this length, the Hall effect is weak in the jet. However experimental evidence shows noticeable differences inside the jet and further investigation is required to understand the dissimilarities between the standard and reverse current cases.

To fully explain the experimental data presented herein we use the PERSEUS[14] code (Plasma as an Extended-mhd Relaxation System using an Efficient Upwind Scheme) that can simulate HED plasmas generally and radial foil dynamics in particular. This code includes the Hall, electron inertia and electron pressure terms, and runs as fast as a standard MHD code by computing the Hall term in a local semi-implicit manner. The electron pressure was "turned off" in these simulations to focus only on the Hall term. The simulations are two-dimensional in r-z cylindrical coordinates. The plasma ionization Z and gas constant γ were assumed constant throughout the computational domain, 3 and 1.15 respectively. Despite these restrictions, simulations confirm the trends observed in both experiments. The ion density for both standard (left) and reverse (right) currents, using a $\log_{10}$ scale in Figure 4-a, shows that the jet with reverse current is taller and denser than the jet with standard currents. The code also reproduces correctly the plasma instabilities visible in in the ablated plasma of Figure 1-a and b. Since the choice of the plasma scale length is rather arbitrary, the ion inertial length criterion of Eq. (2) does not define well the plasma regions where the Hall term dominates. However the simulation gives access to a wealth of plasma parameters and we can compare precisely the Hall electric field with the dynamo electric field to understand the circumstances in which one dominates over the other. We therefore find more judicious to use the Hall-Dynamo Criterion ($C_{HD}$), given by:

$$C_{HD} = \frac{1}{eZn_i} \frac{\|\mathbf{J} \times \mathbf{B}\|}{\|\mathbf{v} \times \mathbf{B}\|} \qquad (4)$$

The $C_{HD}$ compares the strength of the Hall electric field to the strength of the electric field generated by dynamo. Where the ion inertial length criterion requires only the measurements of $n_i$ and Z, $C_{HD}$ also requires the measurements of B, J and v, making it more difficult to determine experimentally. As Figure 4-b shows, Hall electric fields dominate over the dynamo electric field in most of the ablated (outer) plasma. The Hall effect plays a minor role in the remainder of the plasma volume, especially in the plasma jet. This result supports the experimental ion inertial length argument discussed previously. In fact, when no axial magnetic field is present, currents and flows are always perpendicular to the magnetic field in axi-symmetric systems. As a result

$$C_{HD} = \frac{J}{veZn_i}. \qquad (5)$$

We can actually connect both criteria if we use the magnetic field scale length as our characteristic scale length $L$

$$\frac{\delta_i}{L} = M_A C_{HD}. \qquad (6)$$

Since $C_{HD}$ measures the absolute strength of the Hall electric field compared to the dynamo electric field, Eq. (6) shows that the ion inertial length criterion $\delta_i/L$ can overestimate or underestimate the impact of the Hall effect depending if the flow is super-Alfvénic or sub-Alfvénic respectively. Figure 4-c shows that the ion inertial length criterion artificially enhances the impact of the Hall term near the jet axis, where $M_A$ is larger since B is small there, especially in the top section of the jet where most of the volume is devoid of current. It artificially reduces the impact of the Hall effect under the foil, where $M_A$ is smaller due to large B in this region. As a result, the following criterion

$$\frac{\delta_i}{M_A L} \qquad (7)$$

is better apt at judging the importance of the Hall term of plasma dynamics.

In conclusion, experimental data show the Hall term affects the dynamics of strongly collimated plasma jets produced by radial foils, particularly the jet geometry and its density profile. However the Hall criterion $\delta_i/LM_A$ shows that Hall physics dominates only the low-density plasma region surrounding the plasma jet. Surprisingly this effect is strong enough to alter the jet dynamics. The plasma flow stream lines, plotted in Figure 4-a, congregates closer to the axis for reverse electrical currents. This increase in radial inward flow is responsible for the denser, taller jets observed in reverse current cases and it is consistent with the density profiles presented in Figure 3. One can reconcile the impact of the Hall term onto the jet by seeing the ablated plasma surrounding the jet as a virtual electrode where the electric field is dominated by Hall physics. Consequently, Hall-dominated currents and flows in the region surrounding the jet act as boundary conditions to dynamo or Ohmic-dominated currents and flows in the jet region. It is rather evident that HED jets which have $\delta_i/LM_A \gg 1$ and $S \gg 1$ will be strongly influenced by Hall physics. What is more remarkable is that if the ratio of jet density to background density is on the order of 10, a HED jet can have $\delta_i/LM_A \ll 1$ and still be influenced by Hall physics when the background plasma has $\delta_i/LM_A \gg 1$ and $S \gg 1$. Since S is large enough in our experiments to allow the expression of Hall physics, the major obstacle to extend our conclusions to astrophysical jets is in the low Re and Pe of our experimental jets. While they are low compared to astrophysical jets, Re or Pe do not enter directly the GOL scaling of the electric field, only $\delta_i/LM_A$ and S (i.e. $Re_M$) do. As a result, we believe that our experimental and numerical results can be scaled to astrophysical jets when the density ratio of the astrophysical jet density to the stellar background density is smaller than 10. If $\delta_i/LM_A \gg 1$ and S >> 1 in this background plasma, then the electric field surrounding the astrophysical jet will be strongly dominated by Hall physics and the jet will be altered indirectly by Hall physics. If the plasma density of the astrophysical jet is much larger than the background plasma (100 or more) then the conclusion presented herein may not apply. Large Re and Pe can alter the properties of the jet in such ways that the external Hall electric field will not penetrate deep enough in the jet to alter its density profile and experiments working at larger Re and Pe are necessary..

**Acknowledgments**

Research supported by the NSF Grant # PHY-1102471 and the NNSA/DOE Grant Cooperative Agreement # DE-FC52-06NA 00057.

**Figure 1.** Negative black and white Schlieren shadowgraphy for shot # 02580 (standard current) 51 ns into the plasma discharge for shot # 02579 in (reverse current) 53 ns into the plasma discharge. Both shadowgraphs were obtained using a ND:YAG pulsed laser in the green (532 nm). The initial foil location is indicated by the dashed line under which we sketched the 1-mm diameter pin. The direction of plasma instabilities is indicated with arrows.

**Figure 2.** Local electron density of a) shot # 02178 (standard current) 76 ns into the current pulse and b) shot # 02173 (reverse current) 72 ns into the current pulse.

**Figure 3.** Electron density profiles for both standard and reverse currents 1.5 and 2 mm above the foil, normalized to 3 and 2, respectively.

**Figure 4.** a) Plasma ion density, b) Hall-Dynamo Criterion ($C_{HD}$) and c) ion inertial length criterion on the logarithmic scale. All values have been clipped to the minima and maxima of both scales. The white lines in panel a) are plasma flow stream lines.

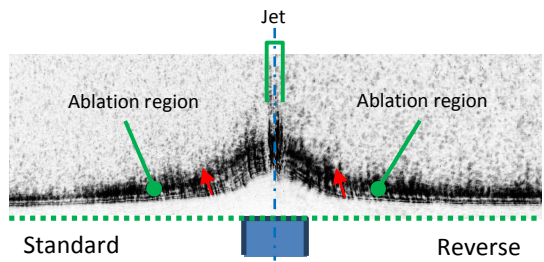

**Figure 1.** Negative black and white Schlieren shadowgraphy for shot # 02580 (standard current) 51 ns into the plasma discharge for shot # 02579 in (reverse current) 53 ns into the plasma discharge. Both shadowgraphs were obtained using a ND:YAG pulsed laser in the green (532 nm). The initial foil location is indicated by the dashed line under which we sketched the 1-mm diameter pin. The direction of plasma instabilities is indicated with arrows.

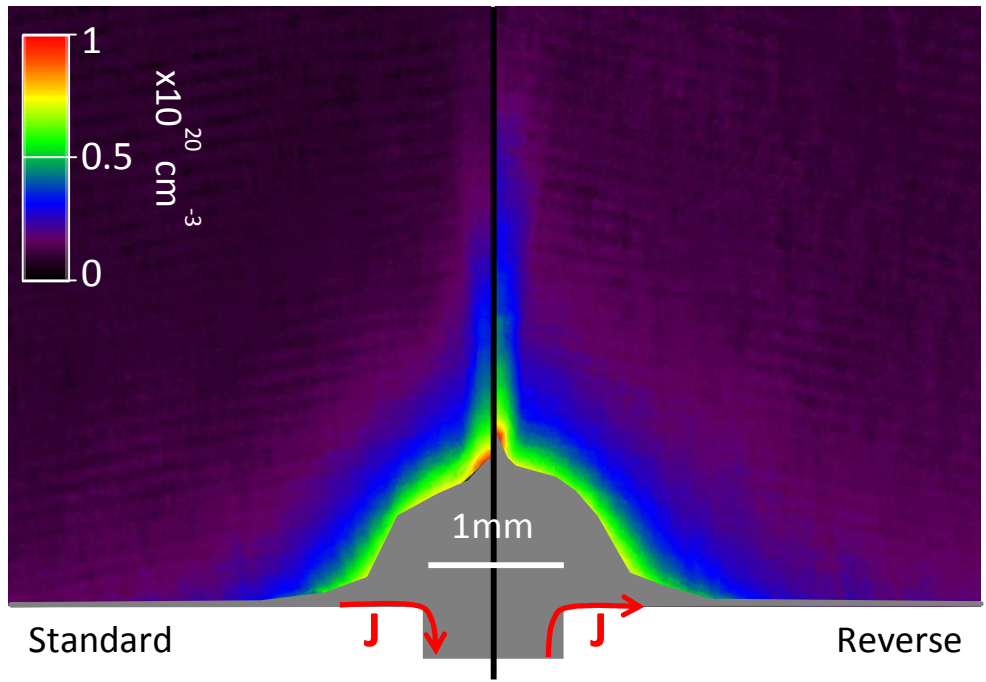

**Figure 2.** Local electron density of a) shot # 02178 (standard current) 76 ns into the current pulse and b) shot # 02173 (reverse current) 72 ns into the current pulse.

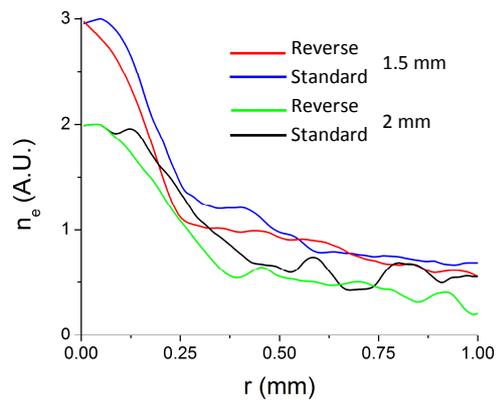

Figure 3. Electron density profiles for both standard and reverse currents 1.5 and 2 mm above the foil, normalized to 3 and 2, respectively.

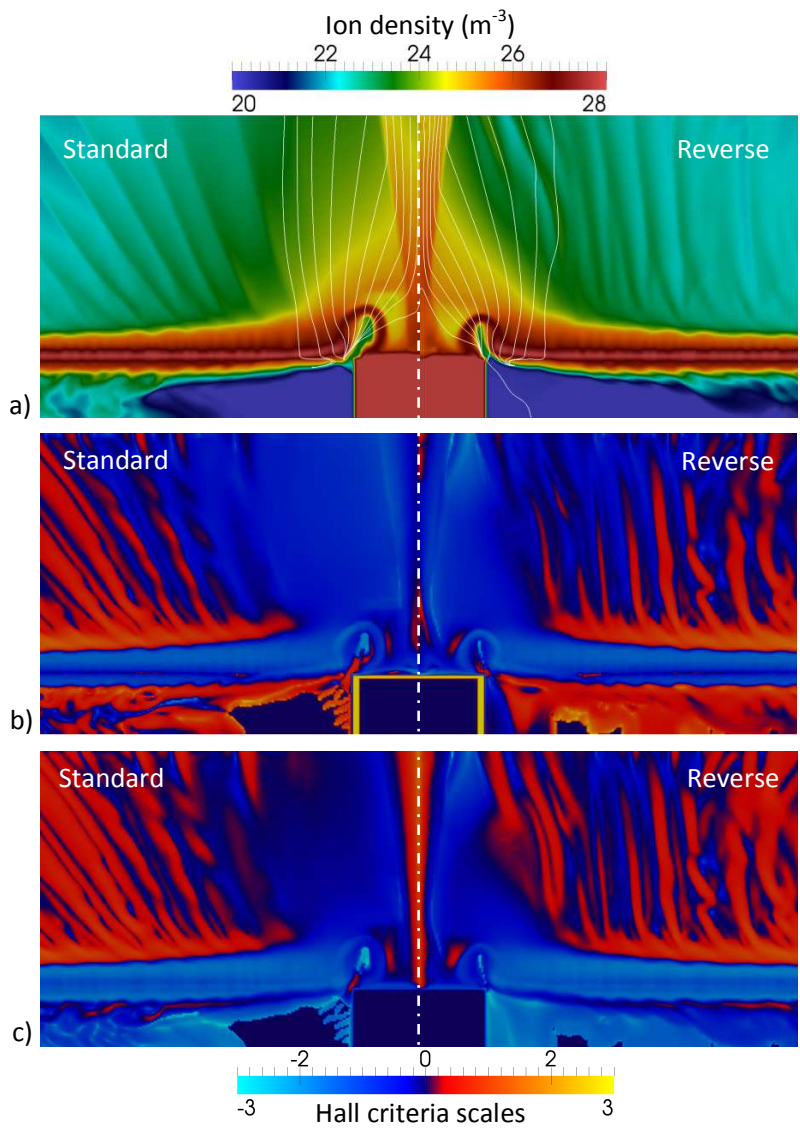

**Figure 4.** a) Plasma ion density, b) Hall-Dynamo Criterion ($C_{HD}$) and c) ion inertial length criterion on the logarithmic scale. All values have been clipped to the minima and maxima of both scales. The white lines in panel a) are plasma flow stream lines.